\documentclass{elsart}
\usepackage{amsmath}
\usepackage{graphics}

\makeatletter

\newcommand{\LyX}{L\kern-.1667em\lower.25em\hbox{Y}\kern-.125emX\spacefactor1000}

\makeatother

\begin{document}

\begin{frontmatter}

\title{A search for energy deposition by neutrinos in matter }

\author{A. Castera, J. Dumarchez, C. Lachaud, F. Vannucci }

\address{LPNHE, Univ. Paris 6\&7, 4 Place Jussieu, Tour 33(RdC), \\
75252 Paris Cedex 05, France. }

\begin{abstract}
An exploratory search for an anomalous energy deposition by neutrinos in a germanium
crystal was performed in the CERN high energy \( \nu _{\mu } \) beam. No signal
was found and a limit is set at a level of about \( 10^{-12} \) of the normal
dE/dx for a minimum ionizing particle. 
\end{abstract}
\end{frontmatter}

\section{Introduction}

We present here the result of a first test looking for energy deposition of
neutrinos in matter, other than via weak interactions. This test consisted in
exposing a high purity germanium crystal to the CERN high energy \( \nu _{\mu } \)
beam. Such a device offers an easy-to-use, high-sensitivity active medium well
suited for an exploratory search. With the very intense neutrino fluxes available
in a beam, it is possible to achieve a meaningful limit even if each neutrino
deposits only a very small amount of energy.  

If detected, neutrino energy losses in matter would have important consequences
in several astrophysical contexts. In particular, an energy loss of neutrinos
in the interior of the Sun could explain the various deficits observed in solar
neutrino experiments. The new Super-Kamiokande measurement of the solar neutrino
spectrum~\cite{Suzuki} excludes a relative energy loss greater than a few per
cent for 10~\( MeV \) neutrinos traversing 600000~\( km \) of matter with
an average density of 1~\( g/cm^{3} \). The result discussed here reaches a
limit lower by two orders of magnitude, when one assumes an energy deposition
proportional to the initial neutrino energy. It is also relevant in the case
of energy transfers in a supernova explosion.

\section{Experimental set-up }

We used a high-purity germanium (HPGe) coaxial diode from Eurisys Mesures\cite{Eurisys}
with an effective volume of 140~\( cm^{3} \) cooled at liquid nitrogen temperature.
Germanium is a semiconductor having an intrinsic band gap of 0.74~\( eV \)
and a mean electron-hole pair creation work of 2.9~\( eV \). This device gives
the lowest energy threshold that can be obtained in an active target of sufficient
mass without requiring heavy cryogenic environment.  

The detector was installed in the CERN high-energy wide band neutrino beam on
a platform between the CHORUS and NOMAD oscillation experiments, 800~\( m \)
away from the primary beryllium target.  

The beam is essentially composed of \( \nu _{\mu } \) of average energy 24~\( GeV \),
with a 1\% contamination of \( \nu _{e} \). It is produced in two spills of
6~\( ms \) duration extracted at the beginning and the end of the 2.2~\( s \)
so-called flat top of the SPS accelerator. The machine cycle is repeated every
14.4~\( s \).   

Synchronisation signals came from the NOMAD experiment, defining 4 gates (we
keep NOMAD's nomenclature):  

-Calib gate, \( 1\, s \) before the first neutrino spill,  

-Nu1 and Nu2 gates, covering the neutrino spills,  

-Mu gate, during the flat-top in between the neutrino spills.  

If the neutrino energy loss was caused by rare but individually detectable effects
(i.e. releasing more than a few \( keV \) per `interaction'), it would be impossible
to disentangle them from ordinary background (such as charged particles correlated
with the neutrino beam or neutrinos weak interactions). Thus, for this first
test, we concentrated on hypothetical processes where neutrinos would each leave
a tiny amount of energy, the neutrino flux resulting in a global increase of
the leakage current during the \( 6\, ms \) accelerator extraction duration.

As a consequence, the read-out was carried out by an infinite time constant
charge-preamplifier, dc-coupled to the HPGe diode. The output of the preamplifier
was sampled and digitized by a Sigma-Delta ADC. On each gate, the Sigma-Delta
loop was reset, and sampling points were accumulated during \( 13.1\, ms \),
constituting a `row'. This arrangement allowed to integrate the energy deposition
during the spill, whilst keeping a sufficient resolution on impulsional energy
deposition (\( 6\, keV \) FWHM on the \( 1.33\, MeV \) \( ^{60}Co \) line),
thus allowing rough spectrometric monitoring and identification of photon or
charged particle interactions in the crystal.

A P-type Ge crystal was chosen, the peripheric Li-doped zone giving some protection
against alphas and low-energy beta and gamma radiations from natural radioactivity.
This protection was enforced by a \( 3\, cm \) thick lead shield surrounding
the detector. The whole set-up was supported by 4 pneumatic insulators. Special
care was taken to enforce ElectroMagnetic Compatibility (EMC).

\section{Analysis }

\begin{figure}[ht]
{\centering \resizebox*{0.9\columnwidth}{!}{\includegraphics{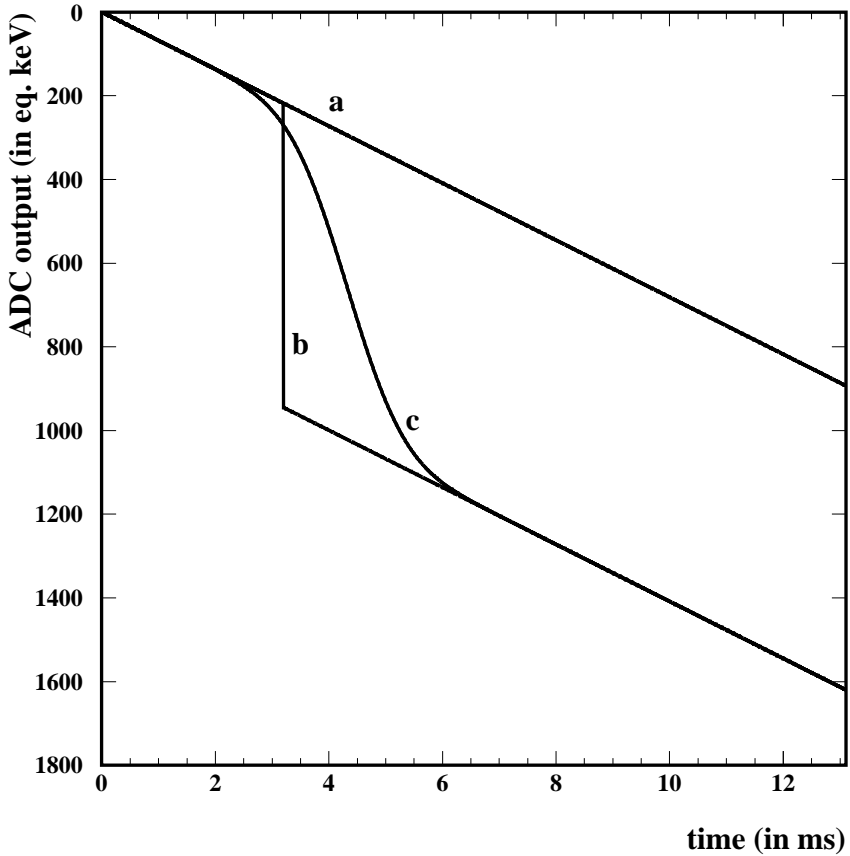}} \par}

\caption{Charge sampling during a 13 ms gate for 3 different cases : in the row (a),
only the leakage current of the Ge crystal is present; row (b) shows the sudden
energy deposition (800 keV in this example) due to a muon across the detector;
row (c) is a simulation of a 800 keV ``signal'' during the neutrino spill.\label{rows}}
\end{figure}

Electron-hole pairs are produced when energy is deposited in the crystal, thus
creating a current through the diode. The leakage current was measured to be
\( 4\, pA \) after a 24-hour warm-up period. When an ionizing particle crosses
the crystal, it gives a fast pulse of \( 1\, \mu s \) duration. Typically,
a minimum ionizing particle crossing the whole thickness of the diode (5~\( cm \))
deposits 30~\( MeV \). As explained in the previous section, the expected signal
behaves very differently : we are looking for a continuous current excess, building
up during the neutrino spill and proportional to the neutrino flux (see Figure~\ref{rows}). 

\begin{figure}[ht]
{\centering \begin{tabular}{cc}
\resizebox*{0.49\columnwidth}{!}{\includegraphics{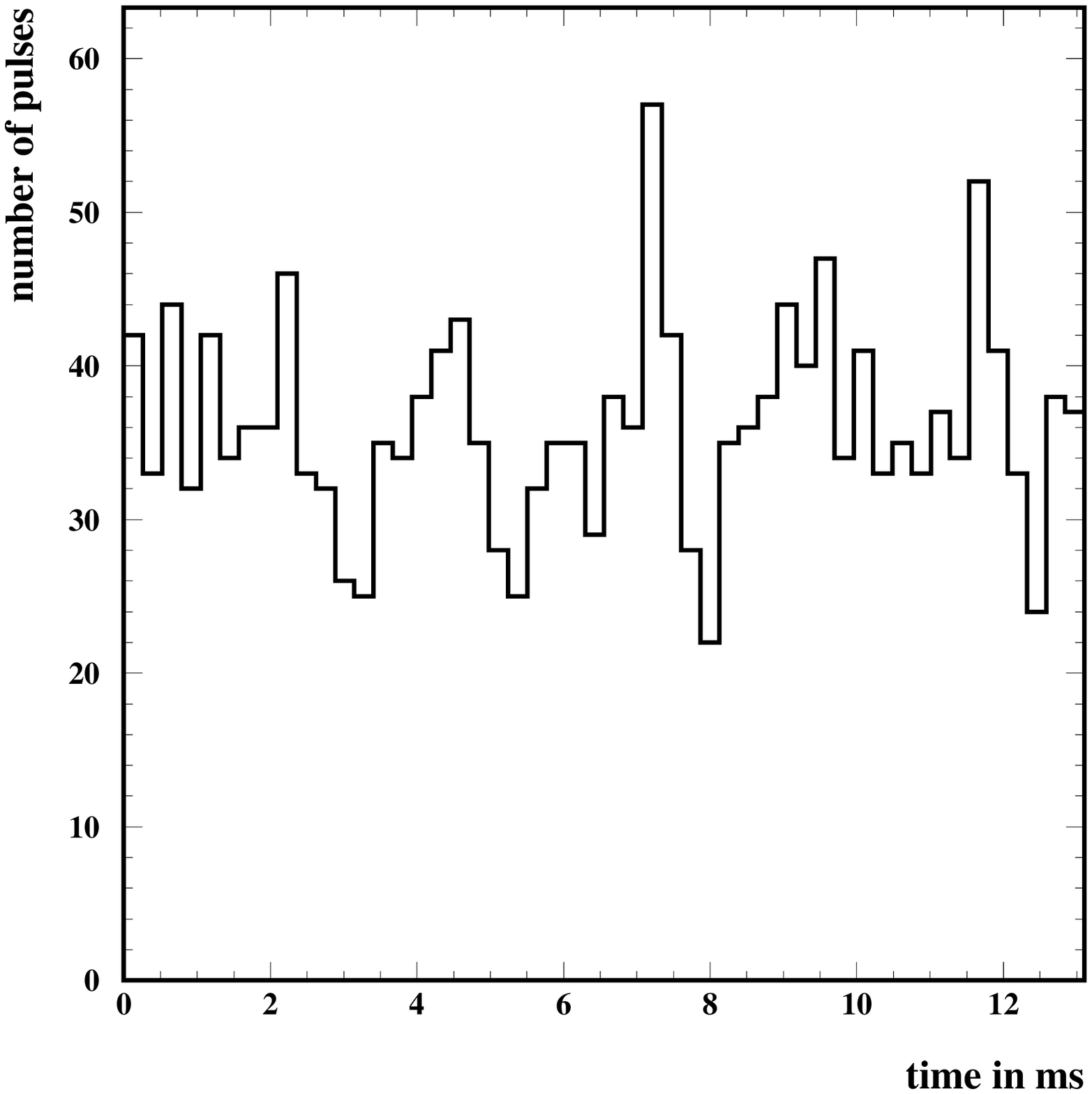}} 
&
\resizebox*{0.49\columnwidth}{!}{\includegraphics{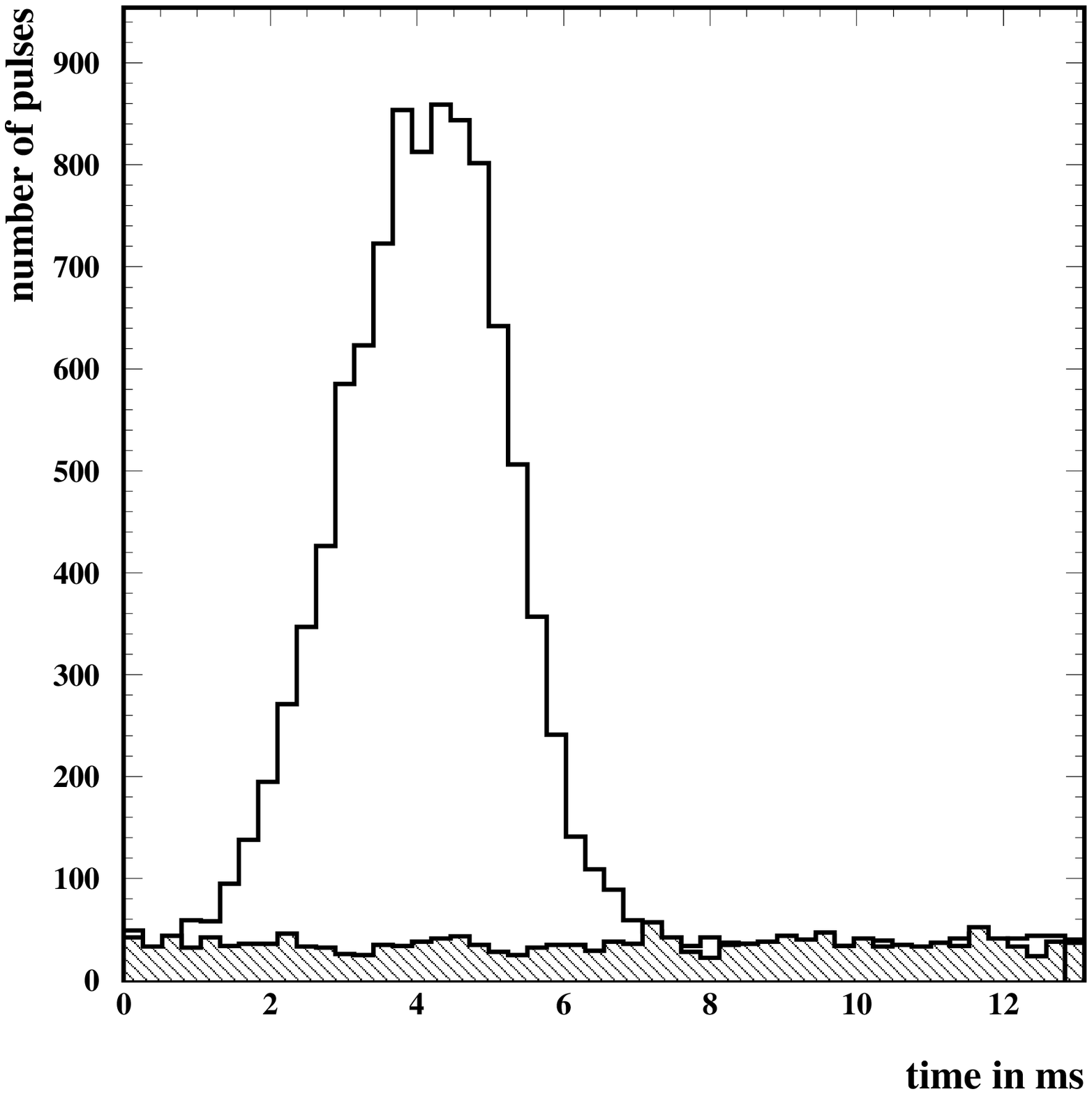}} \\
\end{tabular}\par}

\caption{time distribution (inside the gates) of impulsional energy depositions in the
crystal (above threshold : \protect\( 250\, eV/\mu s\protect \) ; see text)
for the Calib (left) and Nu1 (right) gates. The neutrino pulse time structure
(revealed by the energy deposited by the associated muons) is clearly visible
above a flat distribution, compatible with the one in the Calib gate (reported
on the same plot as a shaded area for the sake of comparison).\label{beamprof}}
\end{figure}

The first step in the analysis is to identify and reject rows contaminated by
photons or charged particles in the crystal. The discriminating cut is set to
\( 250\, eV/\mu s \) over at least two sampling points. This results in rejecting
41\% of the rows in the Nu1 gate, 38\% in the Nu2 gate and 12\% in the Calib
and Mu gates. The time distribution of these rejected energy depositions inside
the \( 13\, ms \) Nu1 or Nu2 gates clearly exhibits the neutrino pulse time
structure (Figure~\ref{beamprof}). 

In the remaining rows, a systematic departure to the straight line is observed
and may be attributed to the periodic reset of the ADC. These systematic effects
prohibit the use of time-correlation related methods. Furthermore, the charge-per-row
versus time distribution shows a semi-periodic pattern over 24~hours, indicating
a perceptible temperature sensibility.

To minimize these systematic effects, two \( 5ms \) long windows are defined
in each row : `in-burst' and `out-burst'. The mean current in each window \( (I_{in},\, I_{out}) \)
is then computed, and the sensitive variable \( S \) is defined as the difference
between the current in- and out-burst \( (S=I_{in}-I_{out}) \).

The next step should be to compare \( S \) in neutrino gates \( (S_{Nu1},\, S_{Nu2}) \)
versus reference gates \( (S_{Calib},\, S_{Mu}) \) as a function of the beam
intensity. The analysis is then restricted to accelerator cycles with all beam
informations available\footnote{
as taken from the NOMAD experiment.
}.

However, as the experiment was performed during a start-up period of the accelerator,
the beam intensity was strongly correlated with time. The risk is then to confuse
beam intensity and time dependant effects. A series of tests using \( S_{Calib} \)
and \( S_{Mu} \) as control samples showed indeed such time dependences (\( >1\sigma  \)).
In spite of a cost in sensitivity and statistics, we have thus decided to use
the difference \( S_{Nu1}-S_{Nu2} \) as a function of the beam intensity difference
\( pot_{Nu1}-pot_{Nu2} \), which is much less time-correlated (\emph{\( pot \)}
stands for Protons On Target)\footnote{
During the start-up period, the neutrino beam intensity changes often and independently
in the two spills as a result of beam steering.
}. In the end only 38\% of the data in the \( Nu1 \) and \( Nu2 \) gates were
used. 

If we write 
\[
S_{Nu1}-S_{Nu2}=A(pot_{Nu1}-pot_{Nu2})+B\]
 the parameter \( A \) is intended to describe the searched-for neutrino-induced
current excess per \( pot \) unit, whilst B regroups all the non beam-correlated
effects. A linear fit on our data gives 

\vspace{0.3cm}
{\centering \begin{tabular}{|c|}
\hline 
\( \widehat{A}=2.5\times 10^{-4}\, keV/10^{10}pot\: ,\: \widehat{\sigma }_{A}\simeq 1.5\times 10^{-3} \)\\
\hline 
\end{tabular}\par}
\vspace{0.3cm}

\begin{figure}[ht]
{\centering \begin{tabular}{cc}
\resizebox*{0.49\columnwidth}{!}{\includegraphics{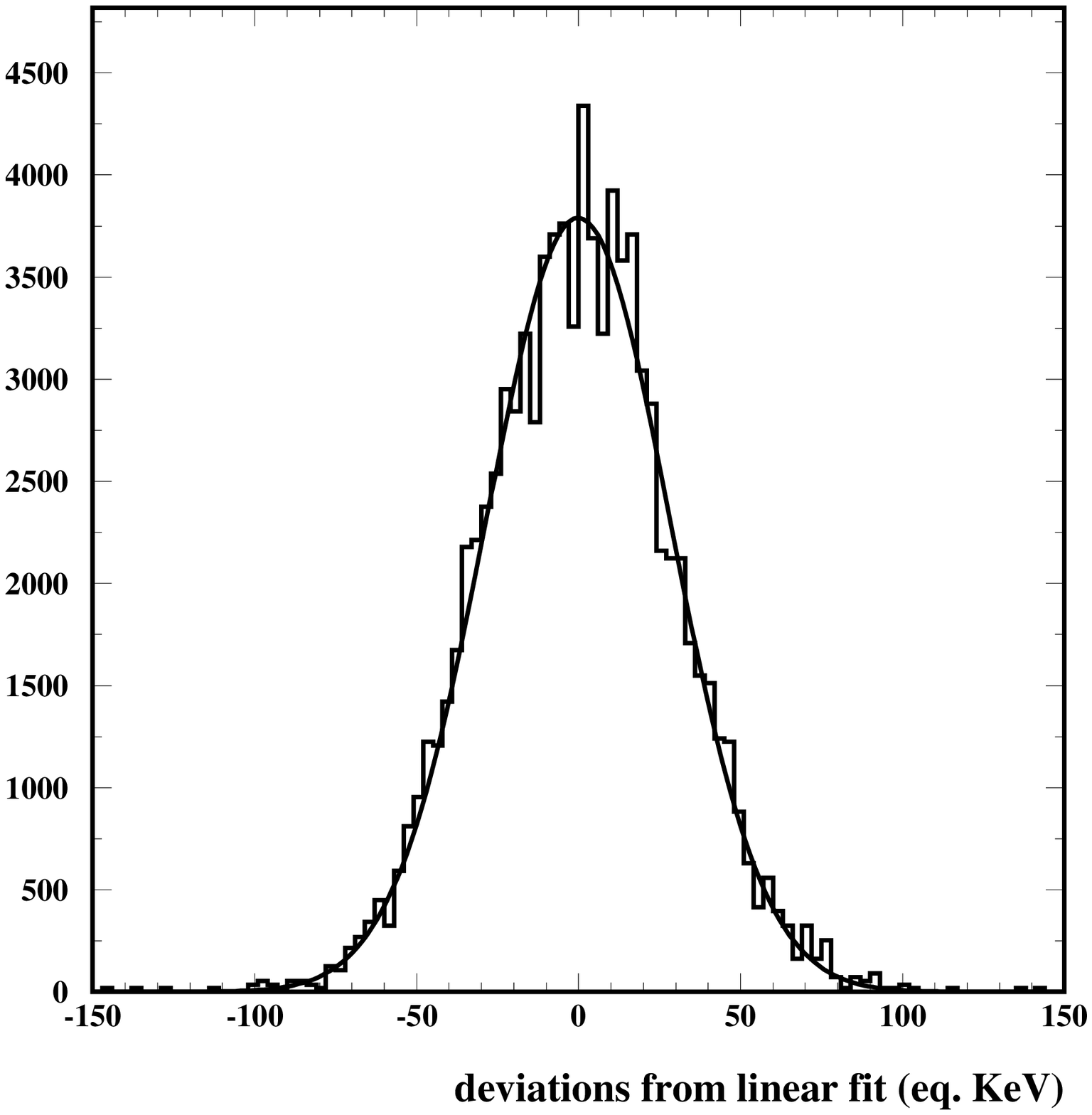}} 
&
\resizebox*{0.49\columnwidth}{!}{\includegraphics{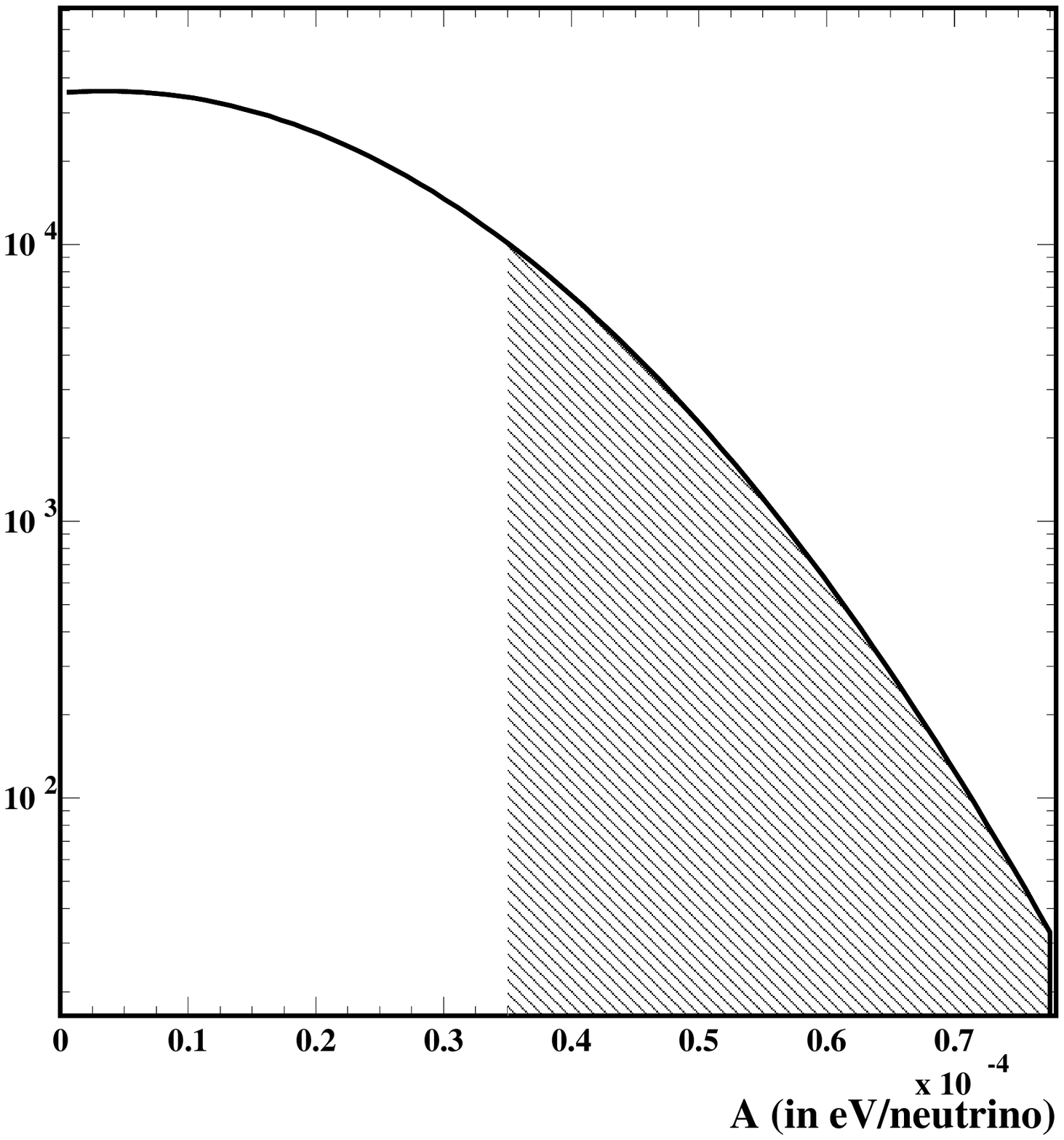}} \\
\end{tabular}\par}

\caption{distribution of the residuals to a linear fit on the current excess in function
of beam intensity (left), and the corresponding \textit{a posteriori} probability
(right) for the energy deposited by a single neutrino across the detector. The
white area covers the 90\% C.L. region.\label{results}}
\end{figure}

Figure~\ref{results}(a) confirms that the residuals to the fit are normally
distributed. Thus under the normal assumption, figure~\ref{results}(b) shows
the conditional distribution of the slope A using the Jeffreys'~prior~\cite{Scherwish}.

This result is compatible with 0 and can be translated into a limit of \( \sim 3\: keV \)
(90\%~C.L.) on the average energy deposited by high-energy neutrinos for a typical
extraction of \( 10^{13}pot \). Such an extraction corresponds to \( 8\times 10^{7} \)
neutrinos of type \( \nu _{\mu } \) and \( \overline{\nu _{\mu }} \) crossing
the crystal. It is then possible to extract a limit on the maximum continuous
energy deposited by a single neutrino.

\section{Conclusion}

We performed a first test to search for a continuous energy loss of neutrinos
in matter. This was done by exposing a high-sensitivity germanium crystal to
the CERN high-energy \( \nu _{\mu } \) beam. It was found that a neutrino of
type \( \nu _{\mu } \) loses less than \( 10^{-5} \)\( eV \) per \( cm \)
of germanium (about \( 10^{-12} \) of the normal dE/dx characterizing a minimum
ionizing particle). Incidently, this would mean an energy loss of \( 10keV \)
for \( \nu _{\mu } \) traversing the whole diameter of the earth; thus an anomalous
energy loss of \( \nu _{\mu } \) cannot explain the deficit of upgoing neutrinos
seen by the SuperKamiokande detector \cite{Fukuda}. For the sake of comparison,
the energy loss due to weak interactions amounts to \( 10^{-2}eV \) per \( cm \)
of germanium for neutrinos of the energy considered here\footnote{
This figure is obtained by assuming that a weak interacting neutrino deposits
all its energy over one ``weak interaction length''.
}. If the loss is predominantly attributed to \( \nu _{e} \)'s (which could
be more probable for an interaction of electromagnetic origin), then the limit
is \( 10^{-3}eV \) per \( cm \) of germanium and per \( \nu _{e} \). This
result applies to a continuous interaction having an energy quantum in the range
1 \( eV \) to 1 \( keV \).

To improve the significance of this result we should carry out the search again
in better experimental conditions (improved mechanical stability, enlarged statistics
...). We are also planning to continue the search at a nuclear reactor where
the neutrino flux is higher and the energies are more relevant for astrophysical
considerations. Finally, we should investigate the use of other devices (bolometers,
RF cavities, etc.) in order to alleviate the limitation on the minimum interaction
quantum, which is intrinsic to the present technique.

\ack{}

We would like to thank P. Repain and L. Serot for their help in the realization
of the experimental set-up. We thank the NOMAD Collaboration for providing the
necessary interfaces with the accelerator informations. Special thanks to the
ICARUS 50l group who kindly shared their already-crowded platform.

\end{document}